# Transport in graphene antidot barriers and tunneling devices

Thomas G. Pedersen[1,2] and Jesper G. Pedersen[1]

[1]*Department of Physics and Nanotechnology, Aalborg University, DK-9220 Aalborg Øst, Denmark*
[2]*Center for Nanostructured Graphene (CNG) , DK-9220 Aalborg Øst, Denmark*

Periodic arrays of antidots, i.e. nanoscale perforations, in graphene enable tight confinement of carriers and efficient transport barriers. Such barriers evade the Klein tunneling mechanism by being of the mass rather than electrostatic type. While all graphene antidot lattices (GALs) may support directional barriers, we show, however, that a full transport gap exists only for certain orientations of the GAL. Moreover, we assess the applicability of gapped graphene and the Dirac continuum approach as simplified models of various antidot structures showing that, in particular, the former is an excellent approximation for transport in GALs supporting a bulk band gap. Finally, the transport properties of a GAL based resonant tunneling diode is analyzed indicating that such advanced graphene based devices may, indeed, be realized using GAL structures.

## 1. Introduction

The excellent transport properties of graphene have been exploited in a first generation of electronic devices based on this material [1]. While these efforts are encouraging, the lack of a band gap means that on-off ratios are rather low in transistors based on pristine graphene, however. Several proposals such as nanoribbons [2] and biased bilayers [3] have been introduced in order to produce a band gap. Recently, also graphene antidot lattices (GALs) [4-12] or nanomeshes [13-17] have been proposed as candidates for semiconducting graphene. These structures consist of holes in the graphene sheet forming a regular antidot lattice/nanomesh (Both names flourish in current literature. Henceforth, in this work, the term *antidot lattice* will be used in accordance with our original proposal [4]). A characteristic of GALs is that they are two-dimensional semiconducting modifications of monolayer graphene that alleviate the need for any external bias. Moreover, GALs circumvent the related issue of Klein tunneling [18] that allows carriers to penetrate electrostatic barriers thereby escaping confinement. In fact, GAL barriers are essentially of the mass (rather than electrostatic) type, which are not subject to Klein tunneling [19]. Hence, antidot barriers enable tight confinement as well as effective transport barriers.

Recently, several experiments on transport in GALs have been reported [9-17]. The structures are produced via lithography and/or etching using various types of masks leading to lattice periods ranging from about 30 nm to several hundred nanometers. Notably, greatly improved on-off ratios have been demonstrated [13,15]. In addition, magnetoresistance oscillations in transport have been observed [9,12]. These reports underline the potential of GALs as truly semiconducting graphene devices. On the theoretical side, several studies support the usefulness of antidot structures for transport devices [20-24]. In Refs. [20] and [21], GALs embedded in nanoribbons have been considered assuming full periodicity and a finite number of perforations, respectively. Interestingly, Ref. [21] finds that transmission is highly sensitive to the number and sideways location of the antidots in the nanoribbon. Antidot lattices with full periodicity perpendicular to the transport direction have been studied in Refs. [22-24]. In Ref. [22], Jippo *et al.* studied the limiting cases, where the number of antidots in the transport direction is either one or infinite. They conclude that a transport gap can be observed for a single line of perforations under certain conditions. Similarly, Gunst *et al.* [23] found that transport through as little as two rows of antidots is essentially completely suppressed in the bulk energy gap range. Moreover, in that work, the effects of increasing the number of antidots along the transport direction was systematically studied. Finally, it has been demonstrated that GALs in a pn-junction geometry may be utilized to realize negative differential conductance [24]. Hence, there is by now ample experimental and theoretical evidence for the usefulness of GALs in electronic transport devices.

Nevertheless, several issues concerning transport in GAL structures remain unresolved. In particular, in the present work, two issues will be thoroughly investigated. Primarily, we wish to quantify and compare the performance of various finite antidot barriers having distinct geometrical properties. For GALs defined by a triangular superlattice, two main categories of structures exist: Regular GALs and rotated GALs (RGALs) [25]. Geometrically, the two types differ by the angle between superlattice vectors and the carbon-carbon bonds. Thus, in GAL structures, vectors connecting neighboring hole centers are parallel to the bonds, whereas, in RGAL structures they are rotated by 30 degrees. Importantly, regular GALs always support a full band gap whereas two-thirds of all RGALs support only directional gaps [25]. By including the RGAL structure in the present work, we wish to clarify to what extent a transport gap may arise for geometries that do not support a bulk band gap. To this end, we compare purely longitudinal transmission to more realistic cases, for which transport is averaged over transverse electron momentum. We verify that only GAL geometries having full band gaps support a transport gap after averaging. A secondary issue that is of great importance for device design is whether simpler model barriers can be applied to predict the behavior of actual antidot geometries. To this end, we compare simulations for antidot devices with simpler models based on either gapped graphene [26-27] or Dirac mass barriers [19]. Moreover, resonant tunneling devices will be considered as an example of realistic device geometries and we assess the applicability of simplified models for simulations of such elaborate structures.

## 2. Methods

Throughout the present work, we model all structures using a tight-binding model identical to that of Ref. [4], i.e. by assuming nearest neighbor interaction with a hopping element of $\gamma = 3.033$ eV and ignoring overlap corrections. The geometries investigated will be based on two representative antidot structures: the {5,1} GAL with a band gap of 0.388 eV and a {6,1} rotated GAL (RGAL) with a vanishing band gap. Here, the {*L*,*R*} notation for GAL structures is applied, where *L* and *R* designate the edge length of the unit cell and hole radius, respectively, both in units of the graphene lattice constant [25]. The band structures and unit cells of the two geometries are illustrated in Fig. 1. For the RGAL structure the vanishing band gap is only found for transport in the $\Gamma \to K$ direction. In contrast, along the $\Gamma \to M$ direction a sizeable band gap of 0.93 eV is found. Hence, the transport gap will be highly direction-dependent. In a realistic setting for transport measurements, however, wide metallic source and drain contact are applied. This situation corresponds to the electron transmission being averaged over all transverse *k*-vectors. Consequently, for the transversally averaged conductance we expect a vanishing transport gap for the RGAL geometry.

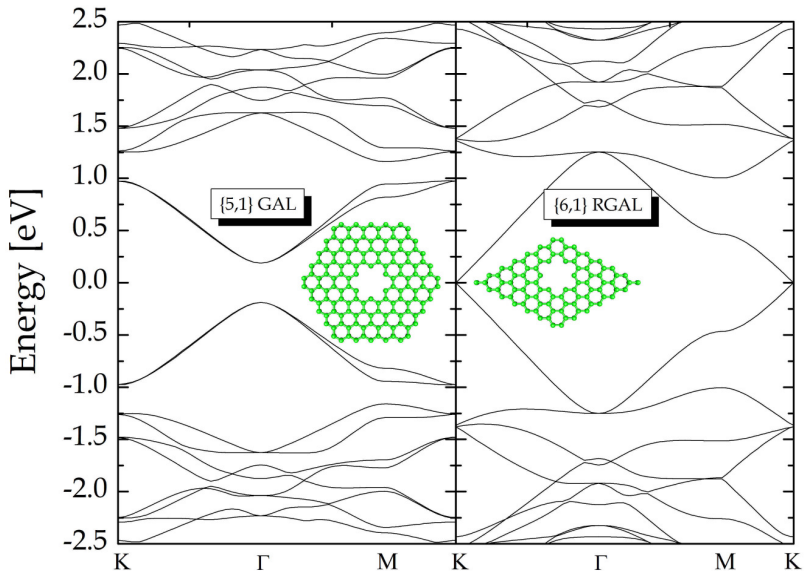

Figure 1. Band structures and unit cells for the {5,1} GAL and {6,1} RGAL structures.

The antidot structures are embedded in either zigzag nanoribbons (ZGNRs) or armchair nanoribbons (AGNRs) such as those shown in Fig. 2. Here, the vertical direction corresponds to longitudinal transport. We always assume periodic boundary conditions in the transverse (horizontal) direction and by choosing the proper nanoribbon width *w*, periodicity for the GAL-based geometries is ensured. Hence, for zigzag embedding as in Figs. 2a and 2b, *N*=10 and *N*=14 ZGNRs are adopted for the {5,1} GAL and {6,1} RGAL, respectively. Similarly, for armchair embedding (Figs. 2c and 2d), *N*=7 and *N*=14 AGNRs are adopted for the {5,1} GAL and {6,1} RGAL, respectively. Coupling between neighbor ribbons include a transverse phase factor $\exp(\pm ik_T w)$, where $k_T$ is the transverse wave vector. We stress that even for purely longitudinal ($k_T = 0$) transport, periodic boundary conditions are applied. Hence, although we consider GALs embedded in nanoribbons, these are only computational devices in the sense that there are no real boundaries. The choice of embedding determines the *k*-space direction of longitudinal transport. Hence, for the cases 2a and 2d, longitudinal transport is along the $\Gamma \to M$ direction while cases 2b and 2c correspond to the $\Gamma \to K$ direction.

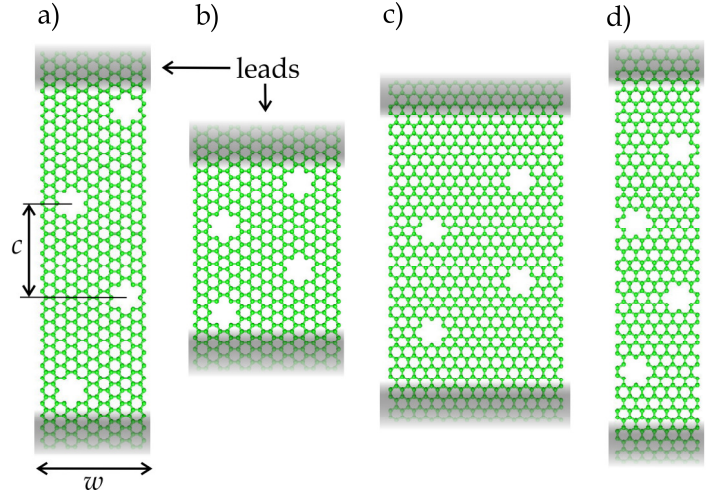

Figure 2. Geometry of antidot barriers: {5,1} GAL in zigzag (a), and armchair (c) embedding, and {6,1} RGAL in zigzag (b), and armchair embedding (d).

For a general geometry, conductance is calculated using the Landauer approach [28]. The leads are assumed semi-infinite and their corresponding surface Green's functions are computed recursively [29, 30]. Coupling the surface Green's functions to the device region (green region in Fig. 2), we find the right (R) and left (L) self-energies $\Sigma_R$ and $\Sigma_L$. Subsequently, the system Green's function is computed as $G = (E - i\varepsilon - H - \Sigma_R - \Sigma_L)^{-1}$, where $H$ is the device Hamiltonian and $i\varepsilon$ is a small imaginary parameter needed for numerical stability for which we take $i\varepsilon = i0.003\,\text{eV}$. Finally, the transmission is found from the usual expression [28]

$$T = \text{Tr}\{\Gamma_R G \Gamma_L G^\dagger\}, \qquad (1)$$

where $\Gamma_{R,L} = i(\Sigma_{R,L} - \Sigma_{R,L}^\dagger)$ are line width functions. For a finite bias, the conductance is obtained by integrating the transmittance $T$ over the transmission window. In the limiting case of a vanishingly small bias, however, the conductance is simply $(2e^2/h)T$.

For geometries supporting a full band gap (rather than a directional one) it is tempting to approximate the actual structure by a simpler barrier. Two such approximations will be investigated here: Gapped graphene [26-27] and Dirac mass barriers [19]. In the gapped graphene model, a staggered on-site potential is introduced in order to produce the band gap. Thus, the on-site energies of orbitals on atoms belonging to the two sub-lattices are shifted up- or downwards by an amount $\Delta$, respectively, leading to a bulk band gap of $2\Delta$. An example of such a gapped graphene barrier is shown in Fig. 3a. Here, the staggered on-site potential is illustrated by the two colors on the sub-lattices. For longitudinal transport, the transverse width $w$ of the gapped graphene nanoribbon is taken to match the full antidot geometry. If transport is averaged over transverse momenta, however, the gapped graphene width can be reduced to, at most, a single hexagon. For a Dirac mass barrier, we assume that the barrier Hamiltonian can be approximated by

$$H = \begin{pmatrix} \Delta(x) & v_F(p_x - ip_y) \\ v_F(p_x + ip_y) & -\Delta(x) \end{pmatrix}, \quad (2)$$

where $v_F$ is the Fermi velocity related to the hopping integral by $v_F = \sqrt{3}a\gamma/2\hbar$ with $a$ the graphene lattice constant. The mass term $\Delta(x)$ is assumed a piecewise constant function that only depends on the longitudinal coordinate $x$. For a barrier of thickness $d$, the mass term is illustrated in Fig. 3b. For a piecewise constant mass term, it is a simple matter of matching wave functions for various regions to find the transmittance. Hence, for a single barrier, the formalism of Ref. [19] can be adopted to show that

$$T(E, k_T) = \frac{E^2 - E_T^2 - \Delta^2}{E^2 - E_T^2 - \Delta^2 \cos^2\left(\frac{\sqrt{E^2 - E_T^2 - \Delta^2}\, d}{\hbar v_F}\right)}. \quad (3)$$

Here, $E_T = \hbar v_F |k_T|$ is the transversal kinetic energy. In the tunneling region $|E| < (\Delta^2 + E_T^2)^{1/2}$, the cosine actually becomes a hyperbolic cosine so that transmission is suppressed exponentially.

a)

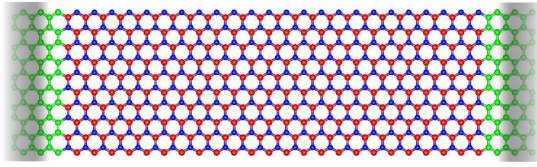

b)

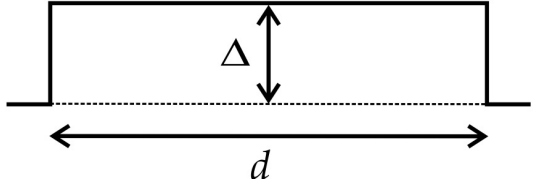

Figure 3. Simplified barriers constructed from (a) gapped graphene with alternating on-site energy and (b) Dirac mass barrier of height $\Delta$ and thickness $d$.

As mentioned above, transmission across a narrow strip of antidots involves electrons with all possible values of the transverse momentum $k_T$. This case corresponds to the realistic experimental scenario of a GAL strip that is narrow in the transport direction but relatively wide along the transversal dimension. In this situation, the experimentally relevant quantity is the transmittance averaged over transverse momenta

$$\bar{T} = \frac{1}{N_k} \sum_{k_T} T(k_T), \tag{4}$$

where $N_k$ is the number of transverse k-points. These are taken from a uniform distribution inside the Brillouin zone $-\pi/w < k_T \leq \pi/w$. Note that the computational cost of averaging the transmission is greatly reduced for gapped graphene compared to the full antidot geometries as a consequence of the much smaller unit cell in the former case. In the Dirac mass barrier model, the transverse period and, hence, Brillouin zone is arbitrary. To compare with results for atomistic models of finite transverse period $w$ we therefore apply the convention

$$\bar{T} = \frac{w}{2\pi} \int_{-\infty}^{\infty} T(k_T) dk_T. \tag{5}$$

Note that enforcing periodic boundary conditions for $k_T = 0$ introduces a doubling of the transmittance. Hence, for these calculations all atomistic tight-binding spectra are divided by a factor of two in order to directly compare with the Dirac mass barrier model.

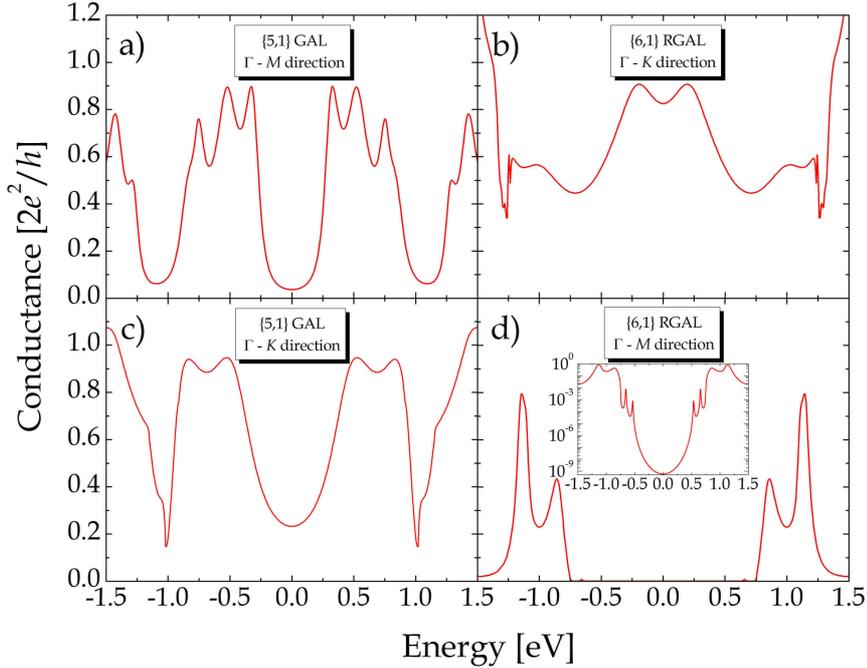

Figure 4. Conductance plots for the four geometries in Fig. 2 assuming purely longitudinal transport. Inset in panel d: Conductance on a logarithmic scale.

## 3. Results for barriers

We start by illustrating the effect of antidot lattice type and orientation on the purely longitudinal transport, i.e. for $k_T = 0$. Figure 4 shows four conductance traces corresponding to the four geometries of Fig. 2. Hence, traces labeled a) to d) correspond to geometries in Fig. 2a to 2d. Obviously, a pronounced effect of both GAL type and orientation is observed. Cases a) and c) display a clear transport gap that agrees with the bulk band gap for the {5,1} GAL of 0.388 eV as expected. The gap is deeper in Fig. 4a due to the thicker barrier. In contrast, the longitudinal transport of the {6,1} RGAL is extremely sensitive to the lattice orientation. Hence, for transport in the $\Gamma \rightarrow K$ orientation, no sign of a transport gap is seen while a huge gap appears for the $\Gamma \rightarrow M$ orientation. The inset in Fig. 4d shows that transmission is suppressed by several orders of magnitude.

Once the transverse average if formed, however, the picture changes completely, as illustrated in Fig. 5. Panels 5a and 5c show that the transport gap survives averaging for the {5,1} GAL irrespective of orientation. Again the depth of the transmission dip is determined by the barrier thickness. For the {6,1} RGAL in Figs. 5c and 5d, on the other hand, the conductance becomes gapless displaying only the linear behavior around the Dirac point reminiscent of pristine graphene. As a consequence, experimental GAL barriers should be fabricated such that antidot superlattice vectors are parallel to carbon-carbon bonds.

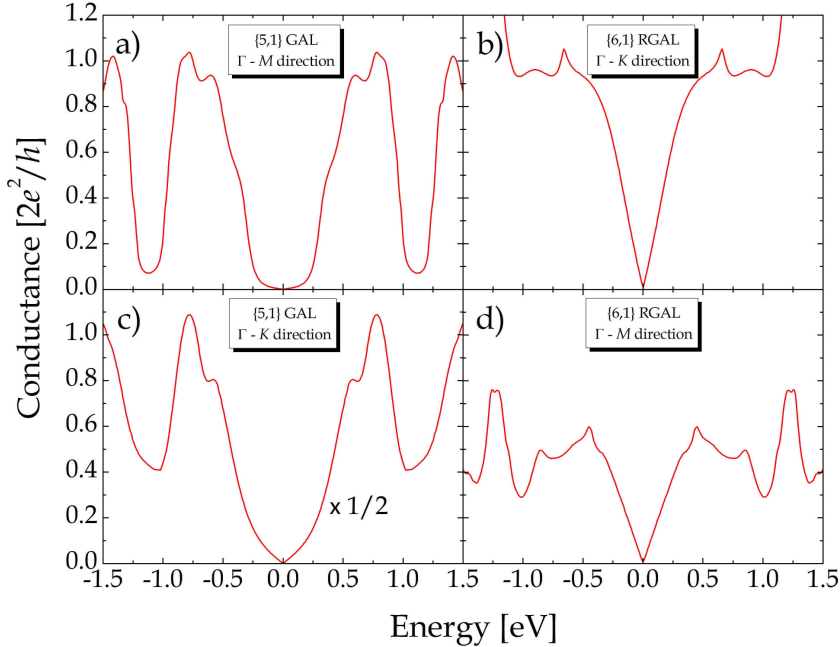

Figure 5. Same as Fig. 4 except that transport is averaged over all possible transverse momenta.

Next, we focus on the comparison of full antidot calculations to those of simplified barriers. Obviously, a successful comparison in the sense of reasonable agreement in the interesting energy range would be important for device simulations. It would conceivably allow for simulations of complicated geometries without excessive computational demands. As for the previous cases, we discuss separately (i) purely longitudinal transport and (ii) transversally averaged transport. The results for the former case are illustrated in Fig. 6. In the plot, results for 2, 4, and 6 antidots are compared to simplified barriers. We have focused on the {5,1} GAL structure with $\Gamma \rightarrow M$ orientation corresponding to the geometry of Fig. 2a. For this case, the longitudinal barrier thickness per antidot is approximately $c = 8a$ and we therefore compare to simplified barriers of thickness 16, 32, and 48 graphene lattice constants *a*. The barrier height $\Delta$ equals half the band gap, i.e. $\Delta = 0.194$ eV, throughout. The effect of increasing barrier thickness is quite clear in all models. Thus, a $32a$ barrier suppresses the transmission by 95% and a near total suppression is seen for a $48a$ barrier corresponding to 6 antidots. In Figs. 6b and 6c, spectra for Dirac mass barriers and gapped graphene are shown, respectively. Overall, the results of the different models are quite similar regarding transmission dip and oscillation period. The magnitudes of transmission resonances differ, though. Thus, whereas the Dirac mass barrier leads to maxima of unit transmittance, c.f. Eq.(3), the gapped graphene model predicts somewhat reduced maxima. The deviation between the models is seen to increase with barrier thickness. As demonstrated in Fig. 6d, this implies that the gapped graphene model is in better quantitative agreement with the full calculation, which also predicts maxima below unity. The overall agreement between all three calculations in the gap range $|E| \lesssim 2\Delta$ is clearly excellent, however. Even the oscillation period is reproduced with remarkable accuracy. Outside the gap range, the details of the true GAL band structure, including additional band gaps near $\pm 1$ eV c.f. Fig. 1, produce notable deviations between full and approximate calculations.

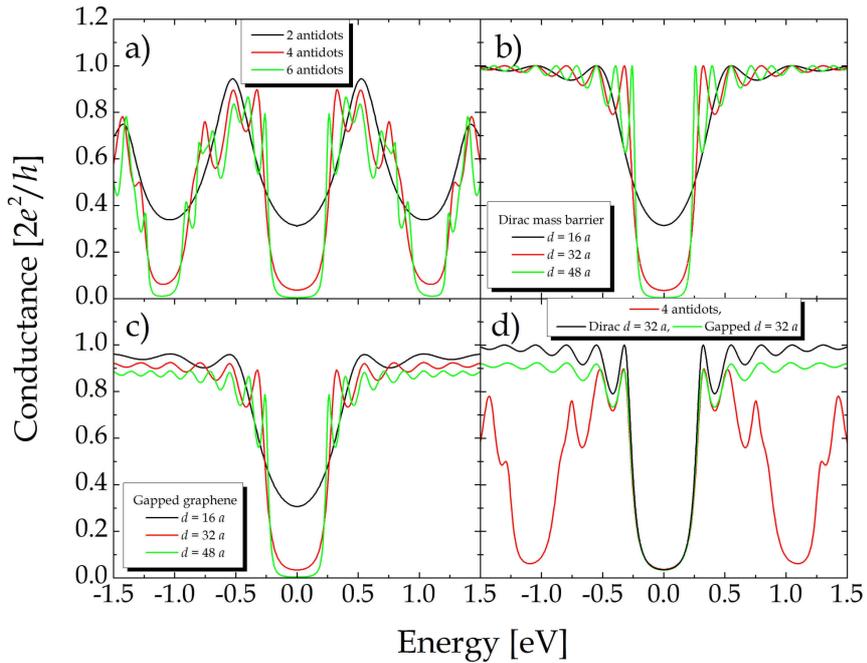

Figure 6. Longitudinal ($k_T = 0$) conductance of {5,1} GAL barriers with zigzag embedding. Full (a) and simplified barriers, i.e. Dirac mass term (b) and gapped graphene (c). Panel d is a comparison of models for 4 antidots and both $d = 32a$ approximations.

In Fig. 7, transversally averaged transport is compared for full and approximate models. Otherwise, all geometries and parameters are identical to those of Fig. 6. Figure 7a shows that increasing the barrier thickness from 4 to 6 antidots only affects the transmission dip marginally upon averaging. This further emphasizes the fact that as little as 4 antidots suffice to efficiently block transport in the gap. However, two lines of these rather small

{5,1} antidots is too little to serve as an efficient barrier. Comparing to the approximations in panels b and c, we see that the Dirac model fails to capture the behavior of the thinnest barriers. This, again, agrees with the finding that two small antidots do not suffice to produce a true mass-type barrier. Otherwise, excellent reproduction of features in the gap region is observed in both approximate models. The direct comparison in Fig. 7d demonstrates excellent quantitative agreement, although the Dirac model tends to slightly overestimate conductance similarly to the finding for longitudinal transport in Fig. 6. We thus conclude that both longitudinal as well as transversally averaged transport through GAL barriers in the vicinity of the fundamental gap can be reliably approximated by these simplified models. Numerically, the gapped graphene model is slightly more accurate but obviously also more computationally demanding than the (nearly) analytical Dirac model.

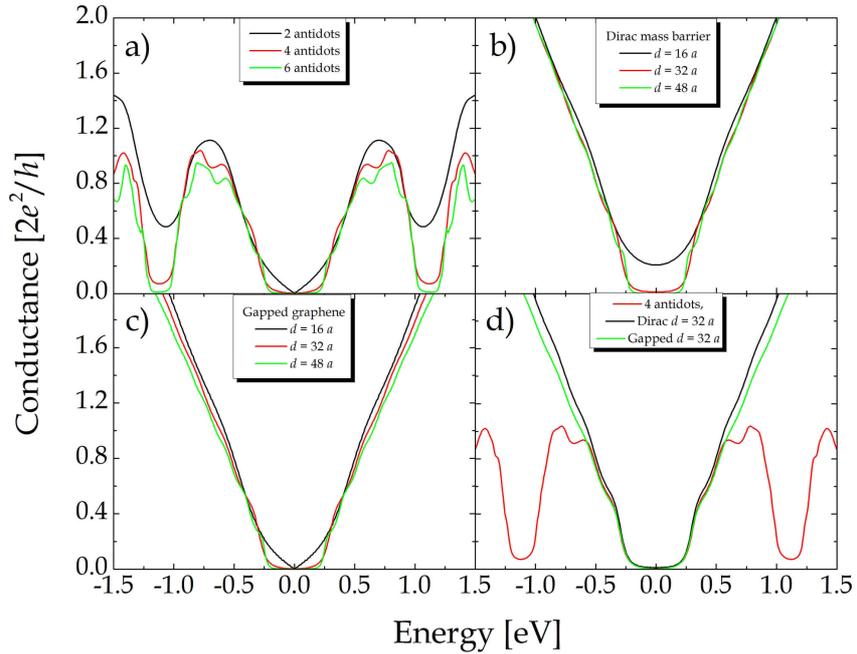

Figure 7. Same as Fig. 6 but after transverse averaging. Note the excellent agreement between different models in panel d.

## 4. Results for resonant tunneling geometry

Individual tunneling barriers such as those considered above could potentially constitute the building blocks for actual devices relying on carrier confinement in graphene. As a prototypical device, we choose here the resonant tunneling geometry containing two barriers separated by a "well" region of intact graphene. Such a structure will lead to discrete transmission resonances in the gap region at certain energies corresponding to standing waves in the well. Hence, highly energy-selective and gate-sensitive transport can be realized in such a structure. The feasibility of this device type has already been studied in the Dirac limit [19]. Here, we study a possible practical realization based on GAL barriers and compare to results for the Dirac limit as well as gapped graphene. We select for illustrative purposes the device geometry in Fig. 8a. Also, simplified models of the Dirac mass type (Fig. 8b) and gapped graphene (not shown) are studied for comparison. For the GAL device, the barrier regions are identical to the barriers in Fig. 2a, i.e. the barrier thickness is $d = 32a$. The well inserted between the barriers is taken to contain 16 rows of intact graphene in the zigzag orientation so that $l = 16a$.

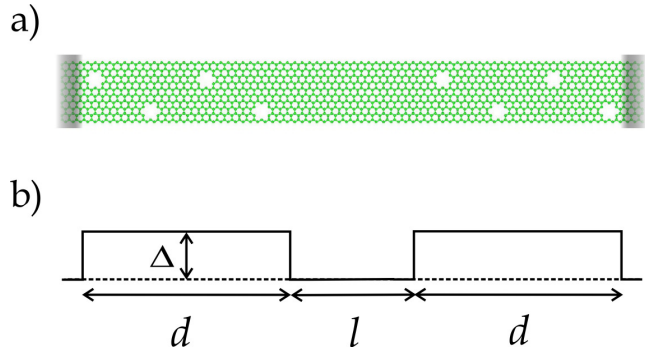

Figure 8. GAL resonant tunneling device. Full atomistic model (a) and Dirac mass barrier approximation (b).

For the resonant tunneling device, we only study longitudinal transport as momentum averaging completely washes out transmission resonances. Presumably, a device for which resonances survive averaging could be constructed by "sandwiching" the tunneling device in Fig. 8a between full barriers. The longitudinal conductance is shown in Fig. 9. Quite distinct resonances at $\pm 0.15$ eV are clearly visible confirming the device functionality. For the present structure, transmittances of nearly 30% are observed and energy-selectivity is demonstrated by the narrow resonance width. When compared to the approximate models, both of these place the resonance slightly too close to the Dirac point. Moreover, the amplitudes are incorrect. Thus, the Dirac model predicts a transmittance of unity precisely at resonance, which is at odds with the full model. This overestimated transmittance is reminiscent of the behavior for individual barriers, c.f. Fig. 6. In contrast, the gapped graphene calculation slightly underestimates the resonance. The discrepancy between resonance energies could be resolved by adopting slightly modified well widths for the simplified models. However, the Dirac model is clearly unable to accurately reproduce the magnitude of the conductance at resonance.

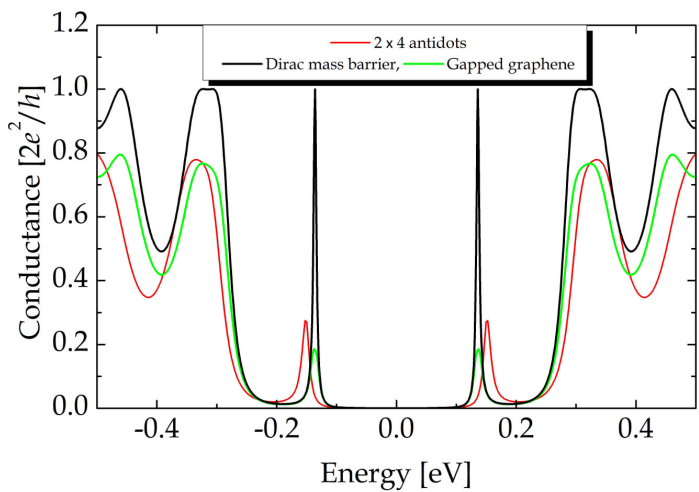

Figure 9. Conductance (longitudinal) of the GAL resonant tunneling diode shown in Fig. 8 and comparison with simplified models.

# 5. Conclusions

In summary, we have analyzed transport through simple GAL barriers and resonant tunneling devices. By comparing geometries with and without a bulk band gap, we show that only the former type supports a full transport gap when all possible transverse momenta are considered. Both types support directional barriers, however. Furthermore, for geometries with a bulk band gap, we find that simple gapped graphene and Dirac mass barriers serve as excellent approximations for the full GAL structure. Thus, simulations of more complicated GAL devices can be performed at greatly reduced computational demand. As an example of a realistic device, a resonant tunneling diode is analyzed. The existence of sharp transmission resonances is established. Again, the simplified gapped graphene model is able to reproduce the main features of the full model.

**Acknowledgments**

Useful discussions with Prof. Mads Brandbyge, DTU Nanotech Denmark, are gratefully acknowledged. The work by J.G.P. is financially supported by the Danish Council for Independent Research, FTP grant numbers 11-105204 and 11-120941. The Center for Nanostructured Graphene (CNG) is sponsored by the Danish National Research Foundation.